\documentclass[showpacs,aps,prb,reprint,superscriptaddress,longbibliography]{revtex4-2}

\usepackage{graphicx}
\usepackage{amsmath,amsfonts,amssymb}
\usepackage{bm}
\usepackage[colorlinks,citecolor=blue,linkcolor=blue,urlcolor=blue]{hyperref}
\usepackage{braket}

\newcommand{\hc}{\text{h.c.}}
\newcommand{\FM}{\text{FM}}
\newcommand{\SC}{\text{SC}}
\newcommand{\cl}{\text{cl}}
\newcommand{\nmc}{\mathcal{N}_{\text{MC}}}

\begin{document}

\def\afflux{\affiliation{Department of Physics and Materials Science, University of Luxembourg, L-1511 Luxembourg, Luxembourg}}
\def\affbra{\affiliation{Instituto de F\'isica, Universidade de S\~ao Paulo, C.P. 66318, 05315–970 S\~ao Paulo, SP, Brazil}}

    \title{Overlap of parafermionic zero modes at a finite distance}

    \author{Raphael L. R. C. Teixeira}\afflux\affbra
    \author{Andreas Haller}\afflux
    \author{Roshni Singh}\afflux
    \author{Amal Mathew}\afflux
    \author{Edvin G. Idrisov}\afflux
    \author{Luis~G.~G.~V. Dias da Silva}\affbra
    \author{Thomas L. Schmidt}\afflux

    \date{ \today }

    \begin{abstract}

   Parafermion bound states (PBSs) are generalizations of Majorana bound states (MBSs) and have been predicted to exist as zero-energy eigenstates in proximitized fractional quantum Hall edge states. Similarly to MBSs, a finite distance between the PBS can split the ground state degeneracy. However, parafermionic modes have a richer exchange statistics than MBSs, so several interaction terms are allowed by the underlying $\mathbb{Z}_{2n}$ symmetry, rendering the effective Hamiltonian governing a pair of PBSs at a finite distance nontrivial. Here, we use a combination of analytical techniques (semiclassical instanton approximation) and numerical techniques (quantum Monte Carlo simulations) to determine the effective coupling Hamiltonian. For this purpose, we go beyond the dilute one-instanton gas approximation and show how finite-size effects can give rise to higher-order parafermion interactions. We find good quantitative agreement between the analytical results and Monte Carlo simulations. We estimate that these finite-size corrections should be observable in some of the recently proposed experiments to observe PBSs in strongly correlated systems.
    \end{abstract}
    \maketitle

    \section{Introduction}
    \label{sec:Intro}

Recent developments in the field of fractional quantum Hall (FQH) systems have allowed the creation of devices containing regions which can be subject to induced superconductivity or interedge scattering~\cite{Lee:ExpSC_QHstates:2017,Cohen:FQHcounterpropagating:2019,Hashisaka:AndreevFQH:2021}. Such systems are theoretically expected to be suitable for hosting bound states with non-Abelian statistics at the interfaces between superconducting and scattering regions~\cite{Barkeshli:TopologicalNematic:2012,Lindner:FractionalizingMajorana:2012,Meng:SCFractional:2012,Clarke:NonAbelianAnyonsFQH:2013,Mong:UniversalTQC:2014,Groenendijk:ParafermionChemicalPotential:2019}. Such states, referred to as $\mathbb{Z}_{n}$ parafermion bound states (PBSs), are generalizations of Majorana bounds states (MBSs), the latter corresponding to PBSs with $n=2$~\cite{Fendley:ParafermionZeroModes:2012}.

Parafermionic states with $n \geq 3$ only appear in strongly correlated systems~\cite{Fidkowski:TopologicalPhases1D:2011,Wen:SPTnonInteracting:2012,Zhang:TRIZ4fractionalJosephson:2014,Chew:FermionizedParafermion:2018,mazza18,calzona18,Khanna:ParafermionMultilegged:2022,Santos:WiresInterface:2017,Nielsen:ReadoutParafermion:2021temp}, which could make realizing PBSs more challenging than realizing MBSs. While it has been suggested that certain one-dimensional systems could host PBSs as well~\cite{Klinovaja:NanowireBundle:2014,Klinovaja:TRIparafermionNanowire:2014,orth15,pedder16,schmidt16,Pedder:MissingShapiro:2017,Teixeira:QD:2021}, FQH edge states are currently considered as the most promising platform to realize a PBS~\cite{Alicea:BluePrints:2016}.

Being one-dimensional and subject to interactions, systems hosting parafermions are usually modeled using bosonization methods \cite{schmidtBosonizationFermionsParafermions2020}. Parafermionic modes emerge at the interfaces between regions where FQH states are coupled by backscattering and superconducting pairing, thus generating nontrivial band gaps. Both effects give rise to cosine terms in the bosonization language, so parafermion systems can be modeled as inhomogeneous sine-Gordon Hamiltonians. While a homogeneous sine-Gordon model is among the rare examples of an exactly solvable interacting model, no such exact solution is known for the inhomogeneous model. Therefore, the theoretical modeling typically rests on approximations.

One such method is the instanton gas approximation, which is based on a semiclassical treatment of the sine-Gordon Hamiltonian~\cite{Coleman:AspectSymmetry,Shifman:abc:1982,Rajaraman:SolitonsInstantons:1982,Chen:TunableSplitting:2016}. Single instantons, time-like kinks of the phase field, are the solutions of the corresponding classical Euler-Lagrange equation and can be used to compute the ground-state energy~\cite{Manton:TopologicalSolitons:2004,Atland:CondensedMatterField:2010}. Multiple instantons can easily be accounted for in the dilute one-instanton gas approximation, where different instantons are assumed to be far apart, so that their interaction can be neglected. This method can be extended towards multiple interacting instantons, also called molecular instantons~\cite{Manton:LagrangianSolitons:1979,bogolmolny1980,ZinnJustin:MultiInstantons:2004}. Used in the context of resurgence theory~\cite{Unsal:ThetaDependence:2012,Misumi:RessurgenceSineGordon:2015}, multi-instantons can offer a path to go beyond the first-order approximation. As we will show, such interacting multi-instanton configurations can become important for PBSs located at a finite distance from each other.

In this work, we go beyond the dilute one-instanton gas approximation~\cite{Chen:TunableSplitting:2016} and include bi-instantons~\cite{Unsal:ThetaDependence:2012}, i.e., configurations including pairs of correlated instantons, in a system that hosts $\mathbb{Z}_{2n}$ parafermion bound states. This second-order process adds an extra ground-state splitting term that exponentially decays twice as fast as the one-instanton term~\cite{Coleman:AspectSymmetry} and contains a distinct oscillatory length dependence in the case of a nonzero chemical potential. These approximate results are then compared to a path-integral quantum Monte Carlo simulation and we find good quantitative agreement.

We proceed to analyze this ground-state splitting correction in terms of higher-order parafermion interactions, i.e., interaction terms containing more than two parafermion operators.
We show that the bi-instanton is associated with a four-parafermion interaction, a result which can be generalized to arbitrary higher orders and associated with $2k$-parafermion interactions.
We briefly discuss the prospects of an experimental observation of bi-instanton corrections.

The paper is organized as follows: in Sec.~\ref{sec:modelmethods} we introduce the sine-Gordon model and the basic assumptions utilized to simplify the action. Moreover, we introduce the effective low-energy model governing the parafermionic modes. In Sec.~\ref{sec:Analytical}, we obtain the finite-size energy splitting of the ground state, first by explaining the dilute one-instanton gas approximation, and then by extending it to include bi-instantons. We continue by comparing the analytical results with Monte Carlo simulations in Sec.~\ref{sec:mc}. Finally, we present our parameter estimates and conclusions in Sec.~\ref{sec:Conclusions}.

 \section{Model} \label{sec:modelmethods}

\begin{figure}[t]
	\begin{center}
		\includegraphics[width=\columnwidth]{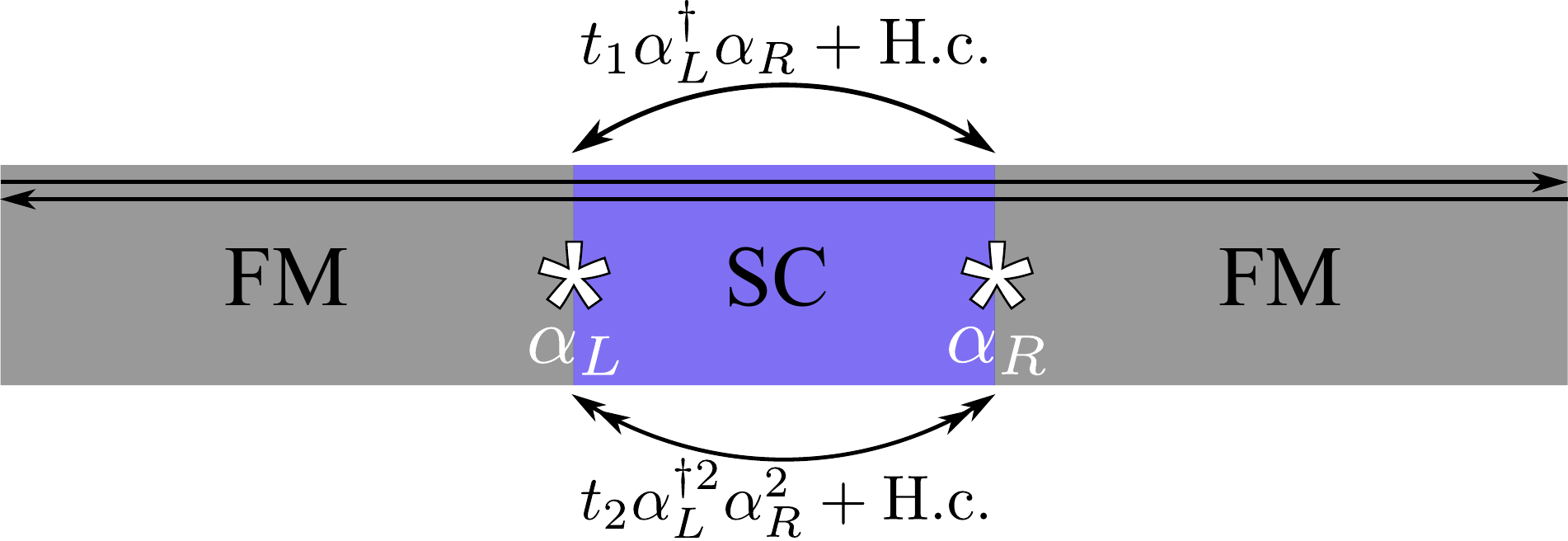}
		\caption{The system comprises a pair of FQH edges with two counter-propagating edge modes at filling factor $1/n$. The FQH edge is subject to induced superconductivity and ferromagnetic coupling leading to an FM-SC-FM junction. The parafermion zero modes appear at the interfaces between the FM and SC regions, which is illustrated by the asterisks.}
		\label{fig:system}
	\end{center}
\end{figure}

The model system for studying parafermion bound states consists of a pair of FQH edges containing two counter-propagating modes. To generate parafermion bound states, it is necessary to engineer two topologically distinct spectral gaps in different regions such that a subgap parafermion bound state appears at the interfaces between these gapped regions~\cite{Barkeshli:TopologicalNematic:2012,Meng:SCFractional:2012}. Many proposals consider an interface between ferromagnetic (FM) and superconducting (SC) regions~\cite{Clarke:NonAbelianAnyonsFQH:2013,Groenendijk:ParafermionChemicalPotential:2019,Chen:TunableSplitting:2016}. Denoting by $\psi_{L,R}(x)$ the annihilation operators for left- and right-moving electrons in the FQH state, the ferromagnet induces a backscattering gap corresponding to a term $\Delta_{\FM}\psi^\dagger_{L}\psi_{R}+\hc$. On the other hand, the proximity effect from a nearby superconductor creates and annihilates Cooper pairs and gives rise to a term $\Delta_{\SC}\psi^\dagger_{L}\psi^\dagger_{R}+\hc$. Hence, a pair of FQH edge states with such an FM-SC-FM junction (see Fig.~\ref{fig:system}) should host parafermion bound states at the interfaces.

\subsection{Sine-Gordon Hamiltonian}

We consider a FQH state with filling factor $1/n$ and group velocity $v$. We first bosonize the left- and right-moving electrons in terms of chiral bosonic fields $\varphi_{L,R}(x)$ such that
\begin{align}
\psi^\dagger_{L,R}(x) = \frac{1}{\sqrt{2\pi n \xi}} e^{-i n \varphi_{L,R}(x)},
\end{align}
where $\xi$ is the correlation length, which is the inverse of the high-energy cutoff, $\xi=v/E_{\rm cutoff}$ (using $\hbar = 1$).

We continue by defining the fields $\phi=(\varphi_{R}+\varphi_{L})/2$ and $\theta=(\varphi_{R}-\varphi_{L})/2$, which satisfy the commutation relations $[\phi(x),\theta(x')]=(i\pi/n)\Theta(x-x')$, where $\Theta$ is the Heaviside function. Hence, $\phi(x)$ and $(n/\pi) \partial_x \theta(x)$ are canonically conjugate variables. The backscattering and pairing terms can then be simplified to $\Delta_{\SC}[\sin(2n\phi)+1]$ and $\Delta_{\FM}[\sin(2n\theta)+1]$, respectively.

The Euclidean action for the whole system thus becomes an inhomogeneous sine-Gordon model,
\begin{align}
	S_{\rm sys}
&=
    \int d\tau dx \bigg\{
        \frac{n v}{2\pi}\left[(\partial_x\theta)^2+(\partial_x\phi)^2\right]+\frac{i n}{\pi}(\partial_x\theta)( \partial_\tau \phi)\notag
	\\
&-\frac{\mu}{\pi}\partial_x\theta
    +\frac{\Delta_{\FM}(x)}{\pi n \xi}\sin(2n\theta)+\frac{\Delta_{\SC}(x)}{\pi n \xi}\sin(2n\phi)\bigg\},
\end{align}
where $\Delta_{\FM}(x)$ and $\Delta_\SC(x)$ vanish, respectively, outside the FM and SC regions and are constant inside those regions. We assume the chemical potential $\mu$ to be constant along the system. We focus on the limit $\Delta_{\FM} L_{\FM}/v \to\infty$, where $L_{\FM}$ is the length of the FM region. In this limit, the field $\theta(x)$ is pinned to a minimum of $\sin(2n\theta)$ inside the FM region, while the field $\phi$ is allowed to fluctuate. By completing the square and considering $\theta(x)$ as constant inside the FM region, the effective action of the system becomes that of a simple sine-Gordon model with a Berry-phase term due to the chemical potential,
\begin{align}\label{eq:action}
	S[\phi]
&=
    \int d\tau \int_{\SC} dx \bigg\{\frac{n}{2\pi v }(\partial_\tau \phi)^2 + \frac{n v}{2\pi}(\partial_x \phi)^2 \notag \\
&\qquad+
    \frac{\Delta}{\pi n \xi}\sin(2n\phi)+\frac{i \mu}{\pi v} \partial_\tau \phi \bigg\}.
\end{align}
Here and in the following, we use $\Delta \equiv \Delta_{\SC}$ and $L \equiv L_{\SC}$ to denote the pairing strength and the length of the superconducting region, respectively. Due to the sine potential, the field $\phi$ is allowed to fluctuate about the minima of the $\sin(2n\phi)$ potential. However, static solutions do not capture the whole picture and instanton solutions that interpolate between different minima are necessary for the ground-state energy. In the next section, we review the dilute one-instanton gas approximation and extend it to include bi-instanton processes.

\subsection{Effective parafermion Hamiltonian}\label{sec:parafermion}

At low energies, the predictions resulting from the sine-Gordon model can be translated into an effective parafermion Hamiltonian, which is valid at energies below the band gaps, $|E| \ll \Delta, \Delta_{\FM}$. To describe the effective parafermion interaction, we consider two parafermion modes described by $\mathbb{Z}_{2n}$ parafermion operators $\alpha_{L, R}$. These operators satisfy the parafermionic commutation relations $\alpha_{L,R}^{2n}=1$, $\alpha_{L,R}^{\dagger}=\alpha_{L,R}^{2n-1}$ and $\alpha_L \alpha_R = e^{i\pi/n} \alpha_R \alpha_L$. A generic interaction Hamiltonian that preserves the $\mathbb{Z}_{2n}$ charge can be written as
\begin{equation}\label{eq:parafermionInt}
	H_{\mathbb{Z}_{2n}}  = \sum_{k = 1}^n H_k,\quad
    H_k = |t_k|e^{i\lambda_k} \alpha_L^{\dagger k}\alpha_R^{\vphantom\dag k} + \hc.
\end{equation}
The phase diagrams of parafermion Hamiltonians, with similar building blocks, have been studied using different methods \cite{Fendley:ParafermionZeroModes:2012,Milsted:CommensurateIncommensurateTQM:2014,iemini17,mahyaehPhaseDiagramMathbbZ2020,woutersPhaseDiagramExtended2022}. This Hamiltonian can be written in the eigenstate basis of the first-order interaction term, $\alpha_L^\dagger\alpha_R^{\vphantom\dag}\ket{q}=-e^{i\pi(q-1/2)/n}\ket{q}$, where the eigenstates $\ket{q}$ are labeled by an integer $q\in \{0,1,2,...,2n-1\}$ which corresponds to the total charge inside the superconducting region in units of the fractional charge $e/n$ \cite{Chen:TunableSplitting:2016}. As a consequence, powers of the first-order interaction term satisfy
\begin{equation}\label{eq:parafermionqbasis}
	\alpha_L^{\dagger k}\alpha_R^{\vphantom\dag k}\ket{q}=(-1)^k e^{i\frac{\pi k}n(q-k/2)}\ket{q},
\end{equation}
such that the energy eigenvalues $E(q)$ of the Hamiltonian $H_{\mathbb{Z}_{2n}}$ can be expressed as a sum,
\begin{equation}\label{eq:EnergyPfk}
	E(q) = \sum_{k = 1}^n 2(-1)^k |t_k|\cos\left(\frac{\pi k}{n}(q-k/2)+\lambda_k\right).
\end{equation}
In the following, we will use a semiclassical instanton calculation as well as a Monte Carlo simulation to determine the effective coupling strengths $t_{1,2}$ and the phases $\lambda_{1,2}$.

\section{Instanton calculation} \label{sec:Analytical}
\subsection{Review of the dilute one-instanton gas}\label{sec:1inst}
In this section, we briefly review the dilute one-instanton gas approximation~\cite{Coleman:AspectSymmetry,Shifman:abc:1982,Rajaraman:SolitonsInstantons:1982,Chen:TunableSplitting:2016} to find the energy splitting between different ground states. In order to obtain the energies, it is necessary to compute the transition rates between different configurations of the field $\phi(x,\tau)$. In imaginary time, such transition amplitudes correspond to the matrix elements,
\begin{align}
    \bra{j_+}e^{-H T}\ket{j_-}
\end{align}
between two stationary states $\ket{j_\pm}$, in which $\phi(x,\tau)$ is pinned at a minimum of the sine potential and which are eigenstates, in the limit $ \Delta L/v \gg 1$, of the Hamiltonian corresponding to the action given by Eq.~\eqref{eq:action}. Moreover, $T\gg\xi /v$ is a large time. This transition amplitude can be conveniently calculated with the action $S[\phi]$.

We focus on classical solutions of the action $S[\phi]$ with constant spatial profile, $\partial_x \phi=0$, \cite{Chen:TunableSplitting:2016} because solutions with a nonzero $\partial_x \phi$ have a larger action and only contribute subleading corrections to the transition amplitude for a given number of instantons. Using this simplification, the equation of motion corresponding to the action~\eqref{eq:action} becomes
\begin{equation}\label{eq:EqMotion}
	\frac{n}{\pi v }\partial^{2}_\tau \phi(\tau)=\frac{2\Delta}{\pi \xi}\cos[2n\phi(\tau)],
\end{equation}
and allows us to define the states $\ket{j}$ which correspond to the stationary solutions $\phi_j=-\pi/4n +j \pi/n$ ($j \in \{0,1,..,2n-1\}$) which minimize the classical action. Solitons correspond to the non-stationary classical solutions and can be found by straightforward integration of Eq.\eqref{eq:EqMotion}. They take the form of a classical field that interpolates between two stationary solutions and is centered at an imaginary time $\tau_0$ (for $\epsilon = \pm$),
\begin{equation}\label{eq:classicalSol}
	\phi_{\cl}^{\epsilon}(\tau)=-\frac{\pi}{4n}+\frac{\pi j}{n} +  \frac{2\epsilon}{n}\arctan\left[e^{\omega(\tau-\tau_0)}\right].
\end{equation}
Here, $\phi_{\cl}^+$ refers to an instanton with final state $\ket{j+1}$, while $\phi_{\cl}^{-}$ denotes an anti-instanton with final state $\ket{j-1}$. Moreover, we defined $\omega=2\sqrt{\Delta v/\xi}$. The action of the soliton field has two different contributions, i.e., a kinetic term ($S_0$) and a Berry-phase term ($S_{Bp}$). While the former is responsible for the system's overall energy scale, the latter will induce an oscillatory behavior in the energy splitting,
\begin{align}\label{eq:classicalaction}
	S[\phi_{\cl}^\epsilon]&= \underbrace{\frac{L n}{\pi v }\int_{0}^{T} d\tau [\partial_\tau \phi_{\cl}^\epsilon(\tau)]^2}_{S_0} +\underbrace{\frac{i L \mu}{\pi v}\int_{0}^{T} d\tau \partial_\tau \phi_{\cl}^\epsilon(\tau)}_{S_{Bp}^\epsilon},	
\end{align}
where we integrated over the position and used Eq.~\eqref{eq:EqMotion} to simplify the action. Calculating the integral for the soliton, one finds
\begin{align}
     S_0 &= \frac{2L\omega}{n\pi v}, \notag \\
    S_{Bp}^\epsilon &= \frac{i \epsilon L\mu}{n v} \equiv i \epsilon \gamma.
\end{align}
As the classical instanton (anti-instanton) interpolates between $\ket{j}$ and $\ket{j+1}$ ($\ket{j-1}$), this allows us to identify the transition rate between these two stationary states. By calculating the path integral with fluctuations $\eta$ around the classical solution, the quantum amplitude of a transition starting in state $\ket{j}$ at imaginary time $\tau=0$ and ending in state $\ket{j+1}$ ($\ket{j-1}$) at $\tau=T$ is given by
\begin{align}\label{eq:GI_calc}
	G&^{\epsilon}=\bra{j+ \epsilon}e^{-H T}\ket{j}=\int \mathcal{D}[\eta]e^{-S[\phi_{\cl}^\epsilon+\eta]}\nonumber\\
	&=T\sqrt{\frac{\pi v S_0}{n}} e^{-S_0 - i \epsilon \gamma}\int\mathcal{D'}[\eta] e^{ -\frac{1}{2}\frac{ n}{\pi v} \iint d\tau  dx \eta \hat{F}^\epsilon \eta}\nonumber\\
	&=T\mathcal{N}\frac{\pi v}{n}\sqrt{S_0}e^{-S_0 - i \epsilon \gamma} {\rm det}'[\hat{F}^{\epsilon}]^{-1/2}.
	\end{align}
where, in the first line, we expanded the action around classical solitons between states $\ket{j}$ and $\ket{j\pm 1}$, such that we have a Gaussian integral over all fluctuations $\eta(x,t)$ which vanish at the boundaries. This remaining path integral contains a translation-invariant zero mode, which can be integrated out and creates a factor $T\sqrt{\pi v S_0/n}$~\cite{Shifman:abc:1982,Coleman:AspectSymmetry}. The remaining integral without zero modes (denoted by the prime in the path-integral measure) is Gaussian, and $\hat{F}^{\epsilon}=-\partial^2_\tau-v^2 \partial^2_x+\omega^2 \sin(2n\phi_{\cl}^{\epsilon})$ is the differential operator associated with the fluctuations. By computing the Gaussian integral in the second line, we obtain the determinant without zero modes, ${\rm det}'[\hat{F}^{\epsilon}]$, multiplied by $\mathcal{N}$, the normalization constant from the measure $\mathcal{D}'[\eta]$~\cite{Coleman:AspectSymmetry}.

Following Refs.~\cite{Coleman:AspectSymmetry,Chen:TunableSplitting:2016}, the determinant can be calculated by multiplying and dividing by $\det[\hat{F}_0]^{-1/2}$, where $\hat{F}_0=-\partial^2_\tau-v^2 \partial^2_x+\omega^2 $ is the differential operator of a harmonic oscillator, and computing the ratio of determinants using the zeta regularization method for a Neumann boundary condition~\cite{Bajnok:FoldedSineGordon:2000},
\begin{equation}
   \frac{\det[\hat{F}_0]^{1/2}}{{\rm det}'[\hat{F}^{\epsilon}]^{1/2}}=\sqrt{\frac{\omega}{L}}.
\end{equation}
As a result, the transition rate becomes
\begin{align}\label{eq:GI}
	G^{\epsilon}=T\mathcal{N}e^{-\omega T /2}\mathcal{K} e^{-S_0 - i \epsilon \gamma},
\end{align}
where $\mathcal{K}=\omega/(\pi\sqrt{n})$. We identify in the results for $G^+$ and $G^{-}$ the transition amplitude arising from an instanton as $[\mathcal{I}]=\mathcal{K}e^{-S_0-i\gamma}$ and for an anti-instanton as $[\bar{\mathcal{I}}]=[\mathcal{I}]^*$. The exponential $e^{-\omega T/2}$ arises from the factor $\det[\hat{F}_0]^{-1/2}$ and is associated with the energy of a harmonic oscillator.

Next, we normalize the stationary states $\ket{j_+}$ and $\ket{j_-}$ with the factor $\sqrt{\mathcal{N}}$, and calculate, within the one-instanton approximation, the most general trajectory between them as that consisting of all possible combinations of well-separated instantons and anti-instantons~\cite{Coleman:AspectSymmetry,Atland:CondensedMatterField:2010},
\begin{align}\label{eq:splitjj1}
&
	\bra{j_+}e^{-H T}\ket{j_-} \notag \\
&=
    e^{-\omega T/2}\sum_{\{n_t\}}\delta_{n_t,j_+-j_-}
	\frac{( T[\mathcal{I}])^{n_{\mathcal{I}}}}{ n_\mathcal{I}!}
	\frac{( T[\bar{\mathcal{I}}])^{ n_{\bar{\mathcal{I}}}}}{n_{\bar{\mathcal{I}}}!}\nonumber\\
&=
	\frac{e^{-\omega T/2}}{2n}\sum_{\{n_t\}}\sum_{q=0}^{2n-1}e^{\frac{i\pi q(j_--j_+)}{n}}
		\frac{( T[\mathcal{I}]e^{\frac{i\pi q}{n}})^{n_{\mathcal{I}}}}{ n_\mathcal{I}!}
		\frac{( T[\bar{\mathcal{I}}]e^{\frac{-i\pi q}{n}})^{ n_{\bar{\mathcal{I}}}}}{n_{\bar{\mathcal{I}}}!}\nonumber\\
&=
    \frac{e^{-\omega T/2}}{2n} \sum_{q=0}^{2n-1} e^{\frac{i\pi q(j_--j_+)}{n}}	e^{2\mathcal{K}Te^{-S_0}\cos(\frac{\pi q}{n}-\gamma)},
\end{align}
where we summed over all combinations of $n_{\mathcal{I}}$ and $n_{\bar{\mathcal{I}}}$ with $n_t=n_{\mathcal{I}}-n_{\bar{\mathcal{I}}}$. In the second line, we used the summation form of the Dirac delta function, expanding $n_t$ and distributing the exponential $e^{\pm i \pi q}$ together with the instanton amplitude.

To obtain an expression for the energy splitting, we expand $\bra{j_+}e^{-H T}\ket{j_-}$ in terms of the eigenstates of a general two-parafermion Hamiltonian and we compare both expressions. The eigenstates $\ket{q}$ of the effective low-energy parafermion Hamiltonian (\ref{eq:parafermionInt}) are similar to Bloch waves and can be written as linear combinations in the $\ket{j}$ basis~\cite{Bajnok:FoldedSineGordon:2000}. By noticing that $\ket{j}=\ket{j+2n}$ due to the periodicity of the sine function, we can write
\begin{equation}
    \ket{q}=\sqrt{\frac{1}{2n}}\sum_{j=0}^{2n-1}e^{-i\pi q j/n}\ket{j}.
\end{equation}
By considering the completeness of the basis $\ket{q}$, we can expand $\ket{j_\pm}$ to obtain
\begin{equation}\label{eq:splitLambda}
	\bra{j_+}e^{-H T}\ket{j_-}=\sum_q \braket{j_+|q}\braket{q|j_-}e^{-E(q)T}.
\end{equation}
Neglecting the constant energy $\omega/2$, we can set Eq.~\eqref{eq:splitLambda} and Eq.~\eqref{eq:splitjj1} equal, to arrive at
\begin{equation}\label{eq:Energy1stO}
	E(q)=-\frac{2\omega e^{-S_{0}}}{\pi\sqrt{n}}\cos\left(\frac{\pi q }{n}-\frac{\mu L}{ n v}\right).
\end{equation}
We note that $E(q)$ decays exponentially as $\propto e^{-L \omega}$, whereas the chemical potential gives rise to a typical oscillation \cite{Chen:TunableSplitting:2016}.
A final equality between Eq.~\eqref{eq:Energy1stO} and Eq.~\eqref{eq:EnergyPfk} relates the parafermion coupling parameters to the microscopic model constants,
\begin{align}
\label{eq:t1}	t_1 = \frac{\omega}{\pi\sqrt{n}}e^{-S_0},\quad
	\lambda_1 =- \frac{\mu L}{ n v} +\frac{\pi}{2n}.
\end{align}
In the dilute one-instanton gas approximation, all higher-order coupling processes are absent, i.e., $t_{k}=0$ for $k>1$.

\subsection{Beyond the dilute one-instanton gas}\label{sec:2inst}

Moving beyond the dilute one-instanton gas approximation, we now consider a system composed of instantons and bi-instantons, the latter corresponding to a correlated two-instanton event~\cite{ZinnJustin:MultiInstantons:2004,Unsal:ThetaDependence:2012,Misumi:RessurgenceSineGordon:2015}. In such a bi-instanton, the change of $\phi(\tau)$ during the transition is still fast compared to the distance between two instantons, but in contrast to the instanton gas limit, this distance is not infinite~\cite{Unsal:ThetaDependence:2012}.

In contrast to a single instanton, the bi-instanton is not an \emph{exact} solution of the classical Euler-Lagrange equation and has a different winding number. Nonetheless, it is a solution up to a correction which is exponentially small in the distance between the instantons, so it can have a significant contribution to the quantum mechanical path integral. Bi-instantons will have an energy scale of the order of $e^{-2S_0}$~\cite{Coleman:AspectSymmetry}, but interactions between the two instantons produce corrections to this energy~\cite{bogolmolny1980}. We consider the action for a configuration of two instantons at positions $\pm \tau_0$ given by (for $\epsilon_1,\epsilon_2 = \pm$)
\begin{align}
	\phi^{\epsilon_1\epsilon_2}(\tau)
&=
    -\frac{\pi}{4n}+\frac{\pi j}{n}  \\
&
    + \frac{2 \epsilon_1}{n}\arctan\left[e^{\omega(\tau+\tau_0)}\right] + \frac{2\epsilon_2}{n}\arctan\left[e^{\omega(\tau-\tau_0)}\right].\notag
\end{align}	
For $\epsilon_1 = \epsilon_2$ such a configuration describes a pair of instantons ($\epsilon_{1,2} = 1$) or anti-instantons ($\epsilon_{1,2} = -1$), whereas for $\epsilon_1 = - \epsilon_2$, the field describes an instanton anti-instanton pair.
The action of such a bi-instanton is twice the classical action of a single instanton plus a positive energy due to the repulsive interaction between them (see the Appendix.~\ref{app:ClassAction2} for a derivation),
\begin{equation}\label{eq:actionBiinst}
	S[\phi^{\epsilon_1\epsilon_2}]=2S_0+4\epsilon_1\epsilon_2 S_0 e^{-2\omega \tau_0}+ i(\epsilon_1+\epsilon_2)\gamma,
\end{equation}
In the following, we focus on the bi-instantons with $\epsilon_1=\epsilon_2\equiv\epsilon$ because these contribute to the energy splitting. In contrast, an instanton--anti-instanton pair ($\epsilon_1=-\epsilon_2$) does not allow additional transitions between different $\ket{j}$, so it just adds a constant value to the energies.

By direct application of the concepts previously introduced for the dilute one-instanton gas, we compute the transition rate due to a single bi-instanton,
\begin{align}\label{eq:GII_full}
	G^{\epsilon\epsilon}
&=
    \bra{j+2\epsilon}e^{-H T}\ket{j}
=
    \int\mathcal{D}[\eta]e^{-S[\phi^{\epsilon\epsilon}+\eta]} \\
&=
    e^{-2S_0- 2i\epsilon \gamma} \int \mathcal{D}[\eta] e^{-4 S_0 e^{-\omega z}}  e^{-\frac{1}{2}\frac{n}{\pi v} \iint d\tau  dx \eta \hat{M}^{\epsilon\epsilon} \eta} \nonumber\\
&=
    T \left(\mathcal{N}e^{-T \omega/2}\right) \mathcal{K}^2 e^{-2S_0- 2i\epsilon\gamma} \int dz e^{-4 S_0 e^{-\omega z}} ,\nonumber
\end{align}
where we followed the same procedure as in Eq.~\eqref{eq:GI_calc}, with the translational invariance of the bi-instanton center of mass now being responsible for the zero mode. Here, $z$ is the distance between the centers of the two instantons and the integral over $z$ comes from the quasi zero modes of $D[\eta]$ \cite{bogolmolny1980}. $\hat{M}^{\epsilon\epsilon}=-\partial^2_\tau-v^2 \partial^2_x+(4\Delta v/\xi)\sin(2n\phi^{\epsilon\epsilon})$ is the differential operator associated with the bi-instanton. The eigenvalues of $\hat{M}^{\epsilon\epsilon}$ are two fold degenerate so $\det'[\hat{M}^{\epsilon\epsilon}]=(\det'[\hat{F}^\epsilon])^2$~\cite{ZinnJustin:MultiInstantons:2004}.

\begin{figure}[t]
    \begin{center}
		 \includegraphics{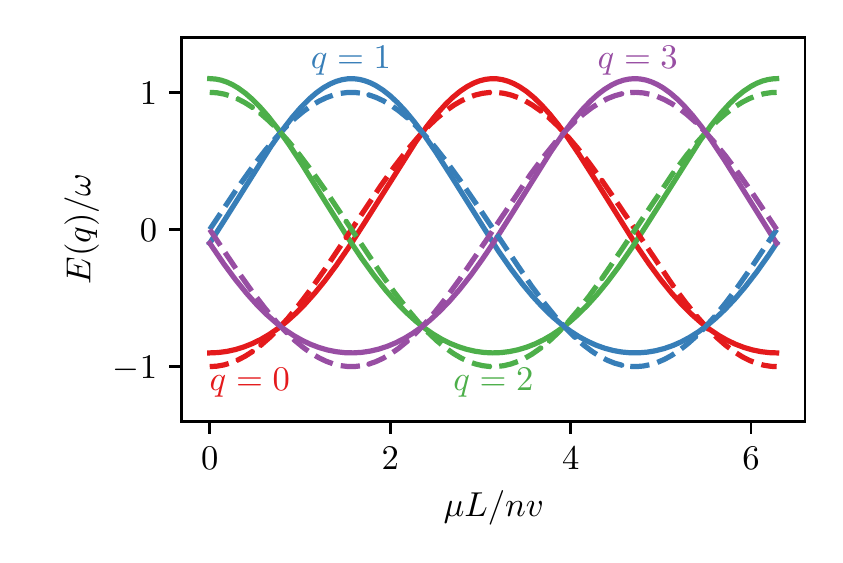}
 		\caption{Energy splitting for ground states with different $q$ as a function of chemical potential. The dashed line corresponds to the one-instanton energy corresponding to only two-parafermion interactions. The solid line corresponds to the energy including four-parafermion terms (bi-instantons corrections), with parameters corresponding to $t_2=0.1 t_1$. The major difference between these two cases is the shifts in the energy crossing at $E=0$.}
 		\label{fig:energyq}
    \end{center}
\end{figure}

The correction to the bi-instanton amplitude due to the instanton interactions, $[\mathcal{II}]=[\mathcal{I}]^2\int dz e^{-4 S_0 e^{-\omega z}}$, can be simplified by using the semiclassical approximation $S_0\gg 1$ and by introducing a regularization parameter $c$~\cite{Misumi:RessurgenceSineGordon:2015} in the instanton-instanton interaction given by Eq.~\eqref{eq:actionBiinst},
\begin{align}\label{eq:instantonInt}
\lim_{c\to0}&\int_{0}^{\infty}dz e^{-4S_0 e^{-\omega z} - c \omega z }  \notag \\
&= \frac{1}{\omega}\lim_{c\to0}\left(\frac{1}{4S_0}\right)^c \int_{0}^{4S_0}ds e^{-s}s^{c-1}\nonumber\\
&\approx\frac{1}{\omega}\lim_{c\to0}\left(\frac{1}{4S_0}\right)^c \Gamma(c) \notag \\
&= - \frac{1}{\omega}\left[\gamma_E+\ln(4S_0)\right] +\mathcal{O}(1/c),
\end{align}
where $\gamma_E\approx 0.5772$ is the Euler number and we need to subtract the divergent term $\mathcal{O}(1/c)$ which corresponds to non interacting instantons. We proceed by generalizing Eq.~\eqref{eq:splitjj1} to a gas consisting of both instantons and bi-instantons, which we call a dilute two-instanton gas approximation. Between two normalized stationary states $\ket{j_+}$ and $\ket{j_-}$, we consider all possible combinations of (anti-)instantons and (anti-)bi-instantons
\begin{align}\label{eq:splitjj2},
&
    \bra{j_+}e^{-H T}\ket{j_-} \notag \\
&=  e^{-\omega T/2 }\sum_{\{n_t\}}\delta_{n_t,j_+-j_-}\nonumber\\
	&\hspace{0.75cm}\times	\frac{( T[\mathcal{I}])^{n_{\mathcal{I}}}}{ n_\mathcal{I}!}
	\frac{( T [\bar{\mathcal{I}}])^{ n_{\bar{\mathcal{I}}}}}{ n_{\bar{\mathcal{I}}}!}
	\frac{( T[\mathcal{II}])^{n_{\mathcal{II}}}}{ n_\mathcal{II}!}
	\frac{( T [\bar{\mathcal{I}}\bar{\mathcal{I}}])^{ n_{\bar{\mathcal{I}}\bar{\mathcal{I}}}}}{ n_{\bar{\mathcal{I}}\bar{\mathcal{I}}}!}\nonumber\\
&=
     \frac{e^{-\omega T/2}}{2n}\sum_{q=0}^{2n-1} e^{\frac{i\pi q(j_- - j_+)}{n}}	e^{2\mathcal{K}Te^{-S_0}\cos(\frac{\pi q}{n}-\gamma)}\nonumber\\
	&\hspace{0.75cm}\times e^{-\frac{2}{\omega}[\gamma_E+\ln(4S_0)] \mathcal{K}^2 Te^{-2S_0}\cos(\frac{2\pi q}{n}-2\gamma)},
\end{align}
where $n_t=n_\mathcal{I}-n_{\bar{\mathcal{I}}}+2n_{\mathcal{II}}-2n_{\bar{\mathcal{I}}\bar{\mathcal{I}}}$ is the net number of instantons. By comparing Eq.~\eqref{eq:splitjj2} with Eq.~\eqref{eq:splitLambda}, it is possible to obtain the energy splitting including the subleading correction due to bi-instantons,
\begin{align}\label{eq:Energy2ndOrder}
	E(q)
&=
    - \frac{2\omega}{\pi\sqrt{n}} e^{-S_0} \cos\left(\frac{\pi q}{n}-\frac{\mu L}{n v}\right) \\
&+
    \frac{2\omega}{\pi^2 n}[\gamma_E+\ln(4 S_0)] e^{-2S_0} \cos\left(\frac{2\pi q}{n}-\frac{2\mu L}{n v}\right)\nonumber.
\end{align}
In Fig.~\ref{fig:energyq}, we show the energy splitting for a $\mathbb{Z}_4$ para\-fer\-mion ($n=2$), comparing the result for the dilute one-instanton gas~\eqref{eq:Energy1stO} (dashed line) with the corrected result including bi-instantons~\eqref{eq:Energy2ndOrder} (solid line) for a value $\sqrt{2} e^{-S_0}=\pi$ and $[\gamma_E+\ln(4 S_0)]/2=0.1$.
While energy crossings between two consecutive $q$ mod 4 occur at the same energy, the crossings between $q$ and $q+2$ mod 4 are shifted.
By setting Eq.~\eqref{eq:Energy2ndOrder} and Eq.~\eqref{eq:EnergyPfk} equal, we recover the leading order result~\eqref{eq:Energy1stO} and, in addition, the second-order coupling amplitude and phase,
\begin{align}
\label{eq:t2}	t_2 &= \frac{\omega}{\pi^2 n}\left[ \gamma_E+\ln(4 S_0) \right]e^{-2S_0},\nonumber\\
	\lambda_2 &=- \frac{2\mu L}{ n v} +\frac{2\pi}{n}.
\end{align}

\section{Monte Carlo} \label{sec:mc}
We start with a discretization of the action~\eqref{eq:action}, using a lattice constant $a_{x}$ and time step $a_\tau$. In the following, we set $v=1$, which leads to
\begin{align}\label{eq:discreteAction}
    S&[\phi]=\frac{n}{2\pi }\sum_{i=1}^{N_\tau}\sum_{j=1}^{N_x}\bigg\{\\
    &\frac{a_x}{a_\tau}(\phi_{\tau_{i+1},x_j}-\phi_{\tau_i,x_j})^2
    +\frac{a_\tau}{a_x}(\phi_{\tau_i,x_{j+1}}-\phi_{\tau_i,x_j})^2\notag\\
&+\frac{\omega^2 a_x a_\tau}{2n^2}\sin(2n\phi_{\tau_i,x_j})+\frac{2i\mu a_x}{n}(\phi_{\tau_{i+1},x_j}-\phi_{\tau_i,x_j})\bigg\},\nonumber
\end{align}
where $\tau_i = i a_\tau$ and $x_j = j a_x$.
Moreover, $N_x$ and $N_\tau$ are chosen such that $N_x a_x = L$ and $N_\tau a_\tau=T$. We choose a rectangular grid, $N_\tau=100$, and set $a_x=a_\tau=a$.

To simulate instanton configurations which connect states $\ket{j_-}$ and $\ket{j_+}$, we assume a twisted boundary condition in the $\tau$ direction, such that $\phi_{\tau_{N_\tau+1},x_i}=\phi_{\tau_1,x_i}+\pi\delta Q/n$~\cite{Bitkevicius:MscThesis}, where $\delta Q = j_+ - j_-$ is an integer. This condition is enforced in the simulation via an energy penalty $\frac{n a_x}{2\pi a_\tau}(\phi_{\tau_{1},x_j}-\phi_{\tau_{N_{\tau}},x_j}+\pi\delta Q/n)^2$, and causes the simulated configuration to have a net number $\delta Q$ of instantons.

The spatial boundary conditions are open, allowing the field to fluctuate freely in the $x$ direction. However, fields which fluctuate in the $x$ direction lead only to subleading contributions compared to configurations which are spatially constant.

The Monte Carlo algorithm is initialized with a vanishing configuration $\phi_{\tau_i,x_j}=0$ for all $i,j$. In each Monte Carlo step, we propose a field update $\phi\rightarrow\phi'$, which is accepted with probability $P=\min(|\exp(-\alpha\Delta S)|,1)$ that is determined by the change of the action $\Delta S = S[\phi']-S[\phi]$.
We introduced an additional parameter $\alpha\geq1$ which will be explained shortly.
Hence, if the updated action $S[\phi']$ is smaller than the original action $S[\phi]$, then $\Delta S<0$ and the update will be accepted ($P=1$). On the other hand, if $\Delta S>0$, we draw a uniformly distributed random number $X\in[0,1]$ and accept the proposed update if $X<P$.
Otherwise, the update is rejected and the original field configuration is kept.

A ``bead update'' which randomly displaces the field along all imaginary-time coordinates has a small acceptance rate, which results in an inefficient simulation.
To achieve an improved convergence towards fluctuations around the classical saddle-point solutions of the path integral, we implemented a local field update scheme at the level of individual nodes, i.e., $\phi'_{\tau_i,x_j} = \phi_{\tau_i,x_j} + \delta\phi$, at a random position $(i,j)$, where $\delta\phi\in[-\delta,\delta]$ is a uniform random displacement and $\delta$ a dynamical interval width.
After 100 Monte Carlo steps, we update $\delta$ depending on the acceptance rate. If the acceptance rate is below $0.6$, we reduce $\delta$ to $0.8\delta$. Typically, this means $\delta$ converges to $0.1$.
Since the action difference after a single local update is independent of the imaginary time $T$ (interpreted as an inverse temperature), an artificial parameter $\alpha$ with $\alpha a_\tau\propto T$ is introduced to obtain a local importance sampling scheme corresponding to an average inverse temperature. We found optimal performance and fast convergence to fluctuations around the classical saddle-point solutions for the choice $\alpha=N_\tau$.

Every $200$ steps, we randomly attempt center-of-mass moves of the field along $\tau$: For a fixed $x_j$, we propose a uniform shift of the field for all $\tau_i$, i.e., $\bm \phi_{x_j}\to\bm\phi_{x_j}+\delta$, where $\bm\phi_{x_j}= (\phi_{\tau_1,x_j},\phi_{\tau_2,x_j}, \ldots)$. The move is accepted according to the same criterion as for the individual node updates. By swapping $\tau$ and $x$, a center-of-mass shift is attempted along the spatial direction. This routine allows slight changes in the center of mass of instantons and improves convergence.

After the system equilibrates (we typically consider $10000$ local updates, although the system usually converges in less than $5000$), we store the final configuration.
The simulation is restarted by creating either an instanton or an anti-instanton, with the same likelihood, at a random imaginary time $\tau$ after which we let the system equilibrate again.
The creation or annihilation of instantons followed by equilibration is repeated $60$ times before the next configuration is stored.
This procedure allows us to sample configurations in the neighborhood of different saddle points which otherwise cannot be reached in reasonable computation time.
For each $\delta Q$, we collect $N_c = 3000$ configurations based on which we compute the observables explained in the following.
To simulate the action, given by Eq.~\eqref{eq:discreteAction}, we use $n=2$, $\omega=\sqrt{2}$, $a=0.6$, and $N_\tau=100$. For each length $N_x$ and chemical potential $\mu$, we simulated configurations with $0\leq\delta Q\leq2$.

    \begin{figure*}[t]
		\includegraphics{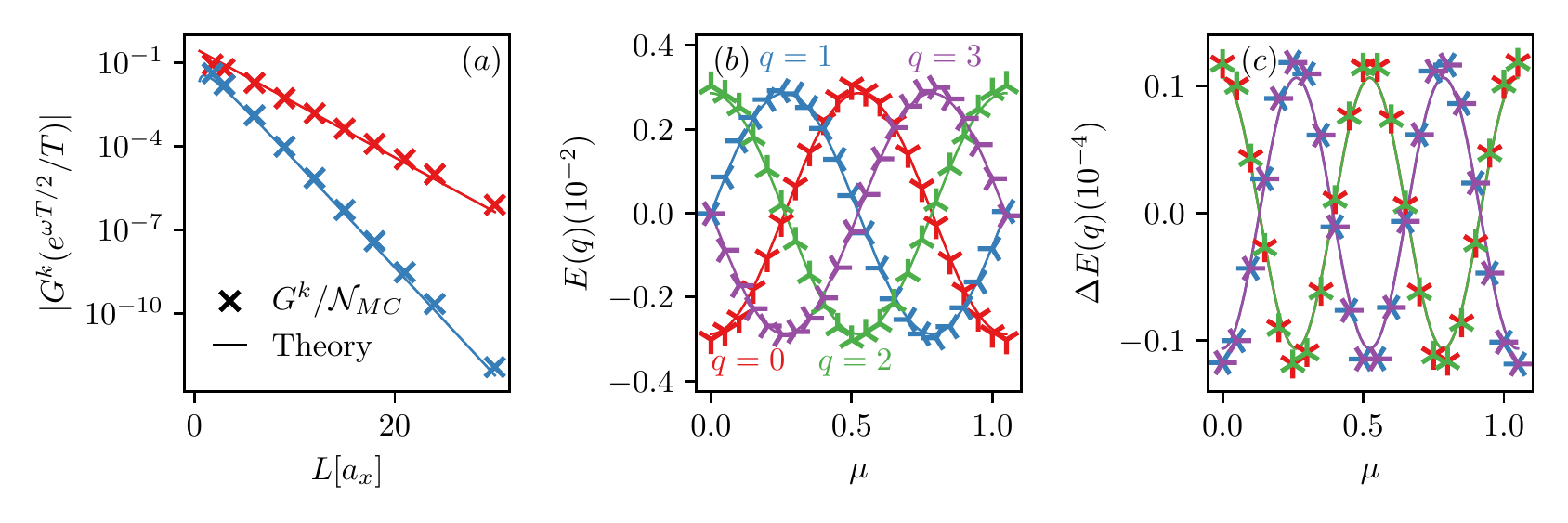}
		\caption{Comparison between Monte Carlo simulations and theory. (a) Transition rates as a function of length $L$ for $\mu=0$ obtained by Monte Carlo (markers) are compared to the theory (solid lines) for $k=1$ (red) and 2 (blue). (b) Energy splitting $E(q)$ as a function of chemical potential for $L=12$. The Monte Carlo simulation, given by Eq.~\eqref{eq:enMC}, matches the theory, given by Eq.~\eqref{eq:Energy2ndOrder}. The energy shift from bi-instantons is not visible as it is small for the chosen parameters. (c) Correction due to bi-instantons (identical for $q=0,2$ and $q =1,3$). The simulation (markers) also agrees with the theory (solid lines).}
		\label{fig:ratepimc}
	\end{figure*}

The twisted boundary conditions effectively decouple different instanton sectors and allow us to simulate quantities such as Eqs.~\eqref{eq:GI_calc} and~\eqref{eq:GII_full}.
Since all $\delta Q$ sectors contribute to physical observables, we define the following average:
\begin{equation}\label{eq:expval}
	\langle A \rangle_{j}=\frac{\sum_{\{\phi_{j}\}} A[\phi_{j}] e^{-S[\phi_{j}]}}{\sum_{\{\phi'_{j}\}} e^{-S[\phi'_{j}]}},
\end{equation}
where $\phi_{\delta Q}$ are configurations containing $\delta Q$ instantons or anti-instantons (negative $\delta Q$). We denote the transition rates of $\delta Q$ instantons as
\begin{equation}
    G^{\delta Q}_{\text{MC}}=\bra{\delta Q}e^{-HT}\ket{0}=\langle e^{-S} \rangle_{\delta Q}.
\end{equation}
The Monte Carlo result depends on a normalization constant $\mathcal{N}_{\text{MC}}$ which is independent of $\delta Q$. To check the results we plot $G^{2}_{\text{MC}}$ and $G^{1}_{\text{MC}}$ [see Fig.~\ref{fig:ratepimc}(a)]. Fitting the expressions for $G^{1(2)}_{\text{MC}}\approx e^{\omega T/2} \nmc G^{1(2)}/T$ we are able to determine $\nmc =  e^{-2\sqrt{2}L/\pi} \sqrt{2}/\pi$, and this result holds for different choices of parameters ($\omega, T$ and $n=2$). Note that even for small lengths, $G^2_{\text{MC}}$ agrees with the analytical result even though it is near the region where the theory is no longer valid. Indeed we later show that the energy splitting deviates from the expected value in this region.

From the transition rates, we follow the derivation of Eq.~\eqref{eq:Energy1stO} to compute the energy splitting. In particular, we note that Eq.~\eqref{eq:splitLambda} can be written as a Fourier transform of transition rates, i.e.
\begin{equation}
	\braket{\delta Q | e^{-HT} | 0} = \sqrt{\frac1{2n}}\sum_{q} e^{-i\pi q\delta Q/n} \braket{q | e^{-E(q)T} | q },
\end{equation}
from which follows
\begin{align}\label{eq:enMC}
    E(q)&\approx-e^{\frac{i\pi q}{n}}\frac{G^1_{\text{MC}}}{\nmc} +e^{\frac{2i\pi q}{n}}\left[\frac{-G^2_{\text{MC}}}{\nmc}+\frac{1}{2}\left(\frac{G^1_{\text{MC}}}{\nmc}\right)^2\right]\notag \\ &+\text{c.c.}.
\end{align}
The energy splitting is plotted in Fig.~\ref{fig:ratepimc}(b) for $L=12$ and shows a good agreement between simulation and theory. We isolate the subleading contributions in Fig.~\ref{fig:ratepimc}(c), which demonstrates that the corrections
\begin{equation}
    \Delta E(q)=e^{\frac{2i\pi q}{n}}\left[\frac{-G^2_{\text{MC}}}{\nmc}+\frac{1}{2}\left(\frac{G^1_{\text{MC}}}{\nmc}\right)^2\right]+\text{c.c}
\end{equation}
are two orders of magnitude lower than $E(q)$ for the chosen parameters, but still in very good agreement with the theory.
\begin{figure}[t]
    \begin{center}
		 \includegraphics{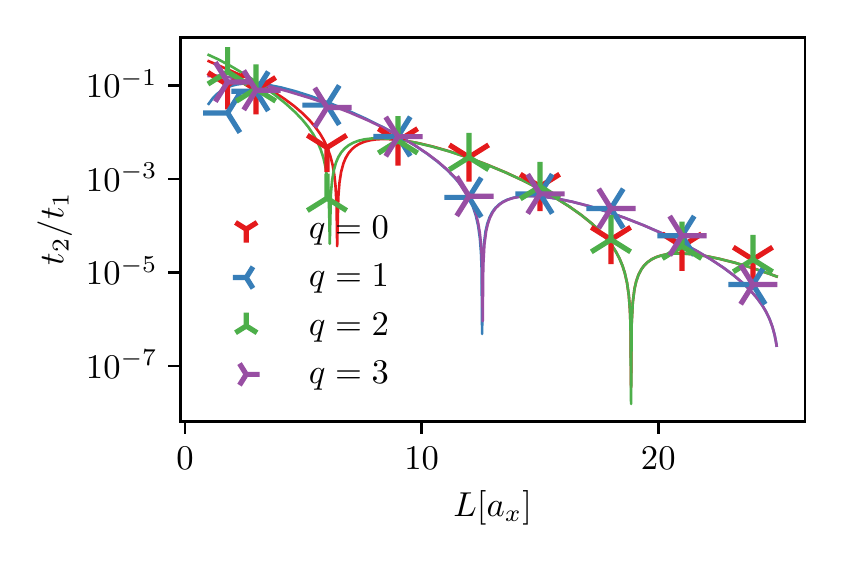}
 		\caption{$|E(q)|$ as a function of length for $\mu=0.5$ with the same color scheme as Fig.~\ref{fig:ratepimc}. Note that for $L<6$, it is possible to clearly observe an asymmetry between the $q=0$ and $2$ sectors. }
 		\label{fig:energyqL}
    \end{center}
\end{figure}
The bi-instanton corrections are more visible when we consider $E(q)$ as a function of length for a fixed chemical potential; see Fig.~\ref{fig:energyqL} for $\mu=0.5$. According to the theory, the level crossing between $q=0$ and $q=2$ at $L\approx6$ is shifted, which we confirm by the simulations. For small lengths, the semiclassical approximation does not fully capture the behavior of the system and we see differences between the analytical results and Monte Carlo simulations. Even so, it is still possible to find regions in which the semiclassical regime describes bi-instantons and the effects are not negligible.

    \begin{figure}[t]
	\begin{center}
		\includegraphics{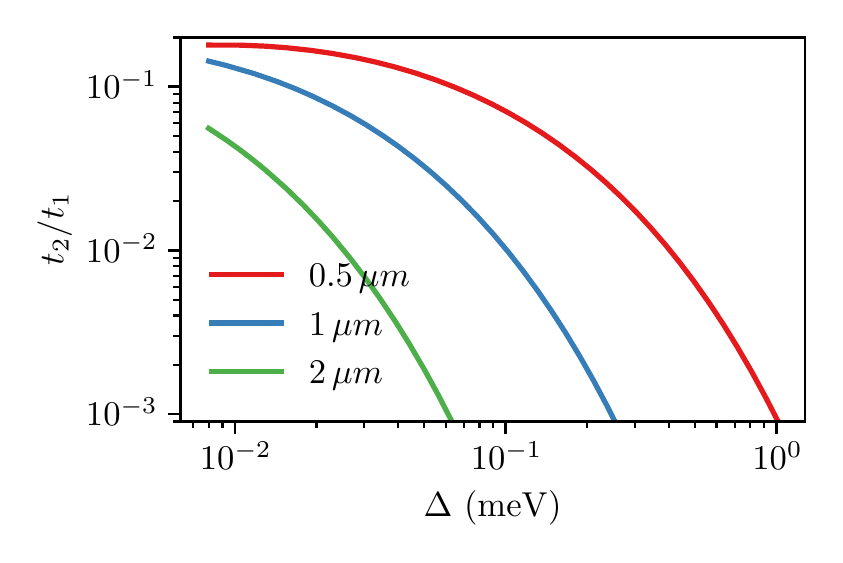}
		\caption{Ratio between energy correction due to one and two instantons, given by Eqs.~\eqref{eq:t1} and ~\eqref{eq:t2}, for typical experimental values $v=10^5$m/s, $E_{\rm cutoff}=2\, {\rm meV}$ and different lengths $L$ over a range of superconducting gap $\Delta$.}
		\label{fig:expvalue}
	\end{center}
\end{figure}

 \section{Conclusion}
 \label{sec:Conclusions}

In this work, we have analyzed the relevance of subleading finite-size effects to the effective description of parafermion bound states existing in a pair of FQH channels with FM-SC-FM interfaces.

We derived a dilute bi-instanton gas approximation and demonstrated that the outcomes are compatible with high-order parafermion interactions in the effective model.
We found a relation between the coupling parameters and the microscopic constants, which reveals that many-body parafermion interaction terms can be observed in small systems and manifest themselves in characteristic oscillations of the ground-state energy level splitting as a function of chemical potential or system size.

We estimate that the subleading correction due to four-parafermion interaction can be of the order of $10\%$ of the leading contribution. Indeed, using typical experimental values for the FQH edges~\cite{Dean:FQHgraphene:2011,Lee:ExpSC_QHstates:2017,Nielsen:ReadoutParafermion:2021temp}, we can estimate $v=10^5$m/s, $E_{\rm cutoff}\sim2\, {\rm meV}$
the bulk gap of FQH, $L\sim \mu m$ and $\Delta \sim 1\, {\rm meV}$,
from which we find that the correction is of the order of $0.1\sim10\%$, as shown in Fig.~\ref{fig:expvalue}.

The bi-instanton approximation is also compatible with Monte Carlo simulations. We highlight that the agreement between analytical and numerical methods indicates the existence of high-order interactions that cannot be explained by previous results. Far from being a mere hindrance, these subleading terms, which do not exist in the case of Majorana bound states, can give rise to novel phases and mechanisms that lie beyond the usual description of parafermion chains.

\acknowledgments

The authors acknowledge fruitful discussions with S.~Groenendijk and M.~Burrello. A.H., E.G.I., and T.L.S. acknowledge support from the National Research Fund Luxembourg under Grants No. C20/MS/14764976/TOPREL and No. C19/MS/13579612/HYBMES. R.L.R.C.T. and L.G.D.S. acknowledge financial support from Brazilian agencies FAPESP (Grants No. 2019/11550-8 and No. 2021/07602-2), Capes, and CNPq (Graduate scholarship program Grant No. 141556/2018-8, and Research Grants No. 308351/2017-7, No. 423137/2018-2, and No. 309789/2020-6).

\appendix

\section{Two-instanton action\label{app:ClassAction2}}
To simplify the calculations~\cite{ZinnJustin:MultiInstantons:2004,Manton:TopologicalSolitons:2004}, we assume one instanton to be centered at imaginary time $\tau_0$ and the other instanton, or anti-instanton, at imaginary time $-\tau_0$, and we choose $j=0$ in Eq.~\eqref{eq:classicalSol}. We also ignore the Berry-phase contribution in this appendix since it is just a boundary term.

The classical solution can be written as
\begin{equation}
	\phi^{+\epsilon_2}(\tau)= -\frac{\pi}{4n}+f_{+}(\tau)+\epsilon_2 f_{-}(\tau),
\end{equation}
where we focus on the case $\epsilon_1 = +$. In the case $\epsilon_2=+$, $\phi^{++}(\tau)$ describes a bi-instanton whereas an instanton--anti-instanton pair corresponds to $\epsilon_2 = -$. The instanton centered at $\pm \tau_0$ is described by $f_\pm (\tau)$,
\begin{equation}
	f_\pm(\tau)=\frac{2}{n} \text{arctan}\left[e^{\omega(\tau\mp\tau_0)}\right].
\end{equation}
The classical action can be divided into a region with positive and negative $\tau$
\begin{align}
	S_+(\phi^{+\epsilon_2})&=\frac{ n L}{\pi v}\int_{0}^{\infty}d\tau\left\{ \frac{1}{2}(\dot{\phi}^{+\epsilon_2})^2+\frac{\omega^2}{4n^2} V(\phi^{+\epsilon_2})\right\}\nonumber\\
	S_-(\phi^{+\epsilon_2})&=\frac{ n L}{\pi v}\int_{-\infty}^{0}d\tau \left\{ \frac{1}{2}(\dot{\phi}^{+\epsilon_2})^2+\frac{\omega^2}{4n^2} V(\phi^{+\epsilon_2})\right\},
\end{align}
where we already integrated over $x$ and assumed $V(\phi)=\sin(2n\phi)+1$. The choice to integrate starting or ending at $\tau=0$ is arbitrary, indeed we could choose any value $\tau_1$ as long as is much smaller than $\tau_0$. Here, we assume $\tau_0\gg1$, in this case $f_{+}$ is small for $\tau<0$ and $f_{-}$ is small for $\tau>0$.

We will focus on $S_+$ expanding it around $f_-=0$, but the same could be done to $S_-$ around $f_+$ and yield an analogous result. Expanding $S_{+}$ around $f_-\ll1$ yields
\begin{align}\
	S_+&(\phi^{+\epsilon_2})=\frac{ n L}{\pi v}\int_{0}^{\infty}d\tau \left\{\frac{1}{2}(\dot{\phi}^{+\epsilon_2})^2+\frac{\omega^2}{4n^2} V(\phi^{+\epsilon_2})\right\}\nonumber\\
	&=\frac{ n L}{\pi v}\int_0^{\infty} d\tau \left\{\left[\frac{1}{2} \dot{f}_+^2 +\frac{\omega^2}{4n^2} V(\frac{\pi}{4n}+f_+)\right]\right.\nonumber\\
	&\hspace{1cm}+\epsilon_2\left[\dot{f}_+\dot{f}_- +\frac{\omega^2}{4n^2} V'(\frac{\pi}{4n}+f_+)f_-\right]\\
	&\hspace{1cm}+\left.\left[\frac{1}{2} \dot{f}_-^2 +\frac{1}{2}\frac{\omega^2}{4n^2} V''(\frac{\pi}{4n}+f_+)f_{-}^2\right]\right\}.\nonumber
\end{align}
The first and third terms are part of the classical action of the instantons centered at $\tau_0$ and $-\tau_0$. This can be seen by rewriting $\frac{1}{2} V''(\frac{\pi}{4n}+f_+)f_{-}^2\approx V(\frac{\pi}{4n}+f_-)$. The linear term in $f_-$ of $S_+$ can be integrated by parts using the equation of motion,
\begin{gather}
	\int_{0}^{\infty}d\tau\left\{\dot{f}_+\dot{f}_- +\frac{\omega^2}{4n^2} V'(\frac{\pi}{4n}+f_+)f_-\right\}\approx \frac{4\omega}{n^2}e^{-2\omega\tau_0}.
\end{gather}
We can do the same procedure for $S_-$, such the classical action of two instantons is
\begin{equation}
	S(\phi^{+\epsilon_2})=2S_0+4\epsilon_2 S_0 e^{-2\omega \tau_0}.
\end{equation}

%

\end{document}